\documentclass[12pt,amsmath]{iopart}

\usepackage{iopams}  
\usepackage{bm}
\usepackage[mathlines]{lineno}

\begin{document}

\title[Contrasting formulations in CPT]{Contrasting  formulations  of  cosmological perturbations  in a  magnetic FLRW cosmology}

\author{ H\'ector Javier  Hort\'ua$^{1,2}$ and Leonardo  Casta\~neda$^{2}$}

\address{$^1$Departamento de F\'isica, Universidad Nacional de Colombia, cra 45 $\#$26-85, Ed. Uriel Gutierr\'ez, Bogot\'a D.C, Colombia}
\address{$^2$Grupo de Gravitaci\'on y Cosmolog\'ia, Observatorio Astron\'omico Nacional,\\
 Universidad Nacional de Colombia, cra 45 $\#$26-85, Ed. Uriel Gutierr\'ez, Bogot\'a D.C, Colombia}
\eads{ \mailto{hjhortuao@unal.edu.co},\mailto{ lcastanedac@unal.edu.co} }
\begin{abstract}
In this paper we contrasted two cosmological perturbation theory formalisms, the \textit{1+3 covariant gauge invariant}   and the  \textit{gauge invariant} by comparing their gauge invariant variables associated with   magnetic field defined in each approach.  
In the first part we give an introduction to each formalism assuming the presence of a magnetic field.    
We found that gauge invariant quantities defined by 1+3 covariant approach  are related with  spatial variations of the magnetic field (defined in the gauge invariant  formalism) between two closed fundamental observers. This relation was computed by choosing the comoving gauge in the gauge invariant approach in a magnetized  universe. Furthermore, we  have derived the gauge transformations  for electromagnetic potentials in the gauge invariant  approach and the Maxwell's equations have been written in terms of these potentials.  
\end{abstract}

\maketitle

\section{Introduction}
Cosmological perturbation theory has become a standard tool in modern cosmology to understand the formation of the large scale structure in the universe, and also to calculate the fluctuations in the Cosmic Microwave Background (CMB)\cite{Padmanabhan}.  The first treatment of perturbation theory within General Relativity was developed by Lifshitz \cite{lifshitz}, where the evolution of structures in a perturbed Friedmann-Lema\^{\i}tre-Robertson-Walker universe (FLRW) under synchronous gauge was addressed. Later, the covariant approach of perturbation theory was formulated by Hawking \cite{Hawking} and followed by Olson \cite{Olson}  where perturbation in the curvature was worked rather than on metric variables. Then,   based on early  works by Gerlach and Sengupta \cite{Gerlach}, Bardeen \cite{bardeen} introduced a full gauge invariant approach to first order in cosmological perturbation theory. In his work,  he built  a set of gauge invariant quantities  related to density perturbations commonly known as Bardeen potentials (see also Kodama \& Sasaki \cite{kodama} for an extensive review). \\

However, alternative representations of  previous formalisms were appearing due to the gauge-problem \cite{sach}.
This issue arises in cosmological perturbation theory due to the fact that splitting  all metric and matter variables into a homogeneous and isotropic space-time plus small desviations of the background, is not unique. Basically, peturbations in any quantity are defined choosing a correspondence between a fiducial background space-time and the physical universe. But, given the general covariance in perturbation theory, which states that there is not a preferred correspondence between these space-times\footnote{The only restriction is that perturbation be small respect to it's value in the background, even so, it doesn't help  to specify the map in a unique way.}, a freedom in the way how to identify  points between two manifolds appears \cite{nakamura1}.  This arbitrariness generates a residual degree of freedom, which would imply that variables might not have a physical interpretation. \\
Following the research  mentioned above, two main formalisms have been developed for studying the evolution of matter variables and to deal with the gauge-problem, that will be reviewed in this paper. The first is known as \textit{1+3 covariant gauge invariant} presented by Ellis \& Bruni \cite{ellis}. This approach is  based on earlier works of Hawking and Stewart \& Walker \cite{steward1}.  The idea is to define covariantly  variables such that they vanish in the background,  therefore,  they can be considered  as  gauge invariant under gauge transformation in according to Stewart-Walker lemma \cite{steward2}. In the  1+3 covariant gauge invariant, gauge-invariant variables  manage the gauge ambiguities and acquire a  physical interpretation.  Since  the covariant variables do not assume  linearization, exact equations are found for their evolution.  The second approach considers  arbitrary order perturbations in a geometrical perspective, it  has been deeply discussed  by Bardeen \cite{bardeen},  Kodama \& Sasaki \cite{kodama},  Mukhanov, Feldman \& Brandenberger \cite{mukhanov},  and  Bruni \cite{bruni1} and it is known as \textit{gauge invariant} approach.  Here, perturbations are descomposed   into the so-called scalar, vector and tensor parts and the gauge invariant are found with  the gauge transformations and  using the Stewart-Walker lemma.  The gauge transformations are  generated by  arbitrary vector fields, defined on the background spacetime and associated with a one-parameter family of  diffeomorphisms. This approach allows to find the conditions for the gauge invariance of any  tensor field,  although at  high  order sometimes appears  unclear. As alternative description of the latter approach, it is important to comment the work done  by Nakamura \cite{nakamura2} where he  splits the metric perturbations   into a gauge invariant and gauge variant part, and  thus, evolution equations are written in terms of  gauge invariant quantities.\\

Given the importance and advantage of these two approaches is nessesary to find equivalences  between them. Some authors have compared  different formalisms, for example  \cite{bruni2} discussed  the invariant quantities found by Bardeen with the ones built on the 1+3 covariant gauge invariant in a specific coordinate system, also the authors in \cite{vitenti}  found  a way to reformulate the Bardeen approach in a  covariant scenario and  the authors in \cite{malik1}  constrasted the non-linear approach described by Malik et al. \cite{malik2} with  the Nakamura's approach.  \\
The purpose of this paper is to present a way for contrasting  the approaches mentioned above. To this aim, we  follow  the methodology  used  by \cite{bruni2}  and \cite{malik3} where a  comparation of   gauge invariant quantities built in each approach is made. However,  we  address  the  treatment in the  cosmological  magnetic fields context,  where cosmological perturbation theory has played an important role for explaining the origin of magnetic fields in galaxies and clusters from a  weak cosmological magnetic field  originated before to recombination era. This means that magnetic fields can leave imprints of theirs influence on evolution of the universe, whether in Nucleosynthesis or  CMB anisotropies \cite{grasso,javier,tina1}.  In fact, the study of primordial magnetic fields will offer a qualitatively  window to the very early universe \cite{giovanini}. 
Cosmological perturbations models permeated by a large-scale primordial magnetic field has been widely worked by Tsagas \cite{tsagasa,tsagasb,barrow} and Ellis \cite{carguese}, where they found the complete equations system  which  shows a direct coupling between the Maxwell and the Einstein fields and also,  gauge invariant for magnetic fields were built  in the  frame of 1+3 covariant approach. Furthermore, in  previous works,  we have obtained a set of equations which describe the evolution of cosmological magnetic fields up to second order in the gauge invariant approach, with their respective gauge transformations for the fields, important for building the gauge invariant magnetic variables \cite{hortua}. Therefore, studying in detail the magnetic gauge invariant  quantities in each one of the formalisms, we can find equivalences between themselves. In addition,  we  have built the invariant gauge for the electromagnetic four-potentials and  the Maxwell equation are written in terms of these potentials.   \\

The outline of the paper is as follows: In section 2 and 3,  the 1+3 covariant and gauge invariant formalisms are reviewed and  the key gauge-invariant variables are defined. In section 4, we introduce the electromagnetic four-potentials in perturbation theory using the gauge invariant formalism, also the gauge transformations are deduced and the Maxwell equations are written here in terms of the  potentials. The section 5 shows the equivalence between the 1+3 covariant and gauge invariant formalism, studying in detail the invariant gauge quantities and discussing the physical meaning of these variables. The last section, is devoted to a discussion of the main results.\\

We use Greek indices $\mu, \nu, ..$ for spacetime coordinates and Roman indices $i, j,..$ for purely spatial coordinates. We also adopt units where the speed of light $c=1$ and a metric signature $(-,+,+,+)$.

\section{The 1+3 Covariant approach: Preliminaries}
We first review the Ellis \& Bruni \cite{ellis} covariant formalism and the  extension of it  with  magnetic field described by  Tsagas \& Barrow \cite{tsagasb,tsagasc} briefly.
The average motion of matter in the universe defines a future-directed timelike  four-velocity  $u^\alpha$, corresponding to a fundamental observer ($u_\alpha u^\alpha=-1$),
and generates a unique splitting of spacetime into the tangent 3-spaces orthogonal to $u_\alpha$.  The second order rank symmetric tensor $h_{\alpha \beta}$ written as
\begin{equation}
h_{\alpha \beta}=g_{\alpha \beta}+u_\alpha u_\beta,
\end{equation}
is the projector tensor which defines the spatial part of the local rest frame of the fundamental observes ($h_{\,\, \alpha}^\beta u_\beta=0$).
The proper time derivative along the fluid-flow lines  and spatial derivative in the local rest frame for any tensorial quantity $T^{\alpha \beta..}_{\quad \gamma \delta..}$ are  given by
   \begin{equation}\label{oper}
\dot{T}^{\alpha \beta..}_{\quad \gamma \delta..}=u^{\lambda}\nabla_{\lambda}T^{\alpha \beta..}_{\quad \gamma \delta..} \quad \textrm{and} \quad  D_{\lambda}T^{\alpha \beta..}_{\quad \gamma \delta..}=h_{\,\, \lambda}^\epsilon h_{\,\, \gamma}^\omega h_{\,\, \delta}^\tau h_{\,\, \mu}^\alpha h_{\,\, \nu}^\beta \nabla_{\epsilon}T^{\mu \nu..}_{\quad \omega \tau..}
\end{equation}
respectively. The operator $D_\lambda$  is the covariant derivative operator orthogonal to $u_\alpha$. The kinematic variables are introduced  by splitting the covariant derivative of $u_\alpha$
into it's spatial and temporal parts, thus
\begin{equation}
\nabla_\alpha u_\beta =\sigma_{\beta \alpha}+\omega_{\beta \alpha}+\frac{\Theta}{3}h_{ \beta \alpha}-a_\beta u_\alpha, 
\end{equation}
where, the variable $a_\alpha=u^\beta\nabla_\beta u_\alpha$ is the acceleration ($a_{\alpha}u^\alpha=0$), $\Theta=\nabla_\alpha u^\alpha$ is the volume expansion, $\sigma_{\beta \alpha}=D_{(\alpha}u_{\beta)}-\frac{\Theta}{3}h_{\beta \alpha}$ is the shear ($\sigma_{\alpha \beta}u^{\alpha}=0, \sigma_{\,\,\alpha} ^\alpha =0$) and   $\omega_{\beta \alpha}=D_{[\alpha}u_{\beta]}$ is the vorticity  ($\omega_{\alpha \beta}u^{\alpha}=0, \omega_{\,\,\alpha} ^\alpha =0$). Also, on using  the totally antisymmetric Levi-Civita tensor $\epsilon_{\alpha \beta \gamma \delta}$, one defines the vorticity vector $\omega^{\alpha}=\frac{1}{2}\omega_{\mu \nu} \epsilon^{\alpha \mu \nu \beta}u_{\beta}$.
A length scale factor $a$ is introduced along the fluid flow of $u_\alpha$ by means of $H=\frac{\dot{a}}{a}=\frac{\Theta}{3}$, with $H$ the local Hubble parameter.   
Now,  we summarize some of results of the covariant studies of electromagnetic fields. The Maxwell's equations in their standard tensor form are written  as
\begin{equation}
\nabla_{[\alpha}F_{\beta \gamma]}=0 \quad \textrm{and} \quad \nabla^{\beta}F_{\alpha \beta}=j_{\alpha}.\label{maxwelldef}
\end{equation}
These equations are  covariantly characterized by the antisymmetric electromagnetic tensor $F_{\alpha \beta}$ and where $j_\alpha$ is the four-current that sources the electromagnetic field \cite{libroellis}. 
Using the four-velocity, the electromagnetic  fields can be expressed as a four-vector electric 
field $E_{\alpha}$ and magnetic field $B_\alpha$ as
\begin{equation}
E_{\alpha}=F_{\alpha \beta}u^{\beta} \quad \textrm{and} \quad B_{\alpha}=\frac{1}{2}\epsilon_{\alpha \beta \gamma \delta}F^{\gamma \delta}u^{\beta}.
\end{equation}
By definition, the electromagnetic four-vectors must be purely spatial and orthogonal to four-velocity ($E_{\alpha}u^{\alpha}=B_\alpha u^{\alpha}=0$). We can  write the electromagnetic   tensor
 in terms of the electric and magnetic fields 
\begin{equation}
F_{\alpha \beta}=u_{\alpha}E_{\beta}-E_{\alpha}u_{\beta}+B^{\gamma}\epsilon_{\alpha \beta \gamma \delta}u^{\delta}.
\end{equation}
The electromagnetic  tensor  determines the energy-momentum tensor of the  field which is given by
\begin{equation}
T_{\alpha \beta}^{(EM)}=-F_{\alpha \gamma}F^{\gamma}_{\beta}-\frac{1}{4}g_{\alpha \beta}F_{\gamma \delta}F^{\gamma \delta}.
\end{equation}
Using the four-vector $u_\alpha$ and the projection tensor  $h_{\alpha \beta}$, one can decompose  the Maxwell's equations (\ref{maxwelldef}) into  a  timelike and a spacelike component,  getting the following set of equations \cite{tsagasc}
\numparts\begin{eqnarray}
h_{\,\,\beta}^{\alpha}\dot{E}^{\beta}=\left(\sigma_{\,\, \beta}^{\alpha}+\omega_{\,\,\beta}^{\alpha}-{\frac{2}{3}}\,\Theta
\delta^{\alpha}_{\,\,\beta}\right)E^{\beta}+ \epsilon^{\alpha \beta \delta \gamma } B_{\delta}\dot{u}_{\beta}{u}_{\gamma}+\mathrm{curl}\,B^{\alpha}-  J^{\alpha}\,,  \\
h_{\,\,\beta}^{\alpha}\dot{B}^{\beta}=\left(\sigma_{\,\, \beta}^{\alpha}+\omega_{\,\,\beta}^{\alpha}-{\frac{2}{3}}\,\Theta
\delta^{\alpha}_{\,\,\beta}\right)B^{\beta}- \epsilon^{\alpha \beta \delta \gamma } E_{\delta}\dot{u}_{\beta}{u}_{\gamma}-\mathrm{curl}\,E^{\alpha} \,,  \label{maxwellcov1}\\
\mathrm{D}^{\alpha}E_{\alpha}= \varrho- 2\omega^{\alpha}B_{\alpha}\,,  \\
\mathrm{D}^{\alpha}B_{\alpha}= 2\omega^{\alpha}E_{\alpha}\,. \label{maxwellcov}
\end{eqnarray}\endnumparts
Where   the curl operator is defined as  $\mathrm{curl}\,E^{\alpha}=\epsilon^{ \beta \alpha \delta \gamma }u_{\delta} \nabla_{\beta}E_{\gamma}$ and the four-current $j_{\alpha}$ splits  along  and orthogonal to $u^{\alpha}$ \cite{tsagasb}, then 
\begin{equation}
 \varrho=-j_{\alpha}u^{\alpha} \quad \textrm{and}\quad J_{\beta}=h_{\,\,\beta}^{\alpha}j_{\alpha}\quad \textrm{with} \quad J_{\alpha}u^{\alpha}=0.
\end{equation}
Finally, using the  antisymmetric  electromagnetic tensor together with Maxwell's equations (\ref{maxwelldef}), one  arrives to a covariant form of the charge density conservation law
\begin{equation}
 \dot{\varrho}=-\Theta\,\varrho-D^{\alpha}J_{\alpha}-\dot{u}^{\alpha}J_{\alpha}.
\end{equation}
In this approach,   Ellis \&  Bruni \cite{ellis} built gauge invariant quantities associated with the orthogonal spatial gradients of the energy density $\mu$, pressure $P$ and fluid expansion $\Theta$. Assuming that the unperturbed background universe is represented by a FLRW metric, the following basic variables are considered
\begin{equation}
 X_{\alpha}=\kappa\, h_{\,\,\alpha}^{\beta}\nabla_\beta \mu, \quad  Y_{\alpha}=\kappa\, h_{\,\,\alpha}^{\beta}\nabla_\beta P \quad \textrm{and}\quad Z_{\alpha}=\kappa\, h_{\,\,\alpha}^{\beta}\nabla_\beta \Theta, \label{GI1}
\end{equation}
where $\kappa=8\pi G$. In fact, the variables such as pressure or energy density are usually nonzero in the FLRW background and so are not gauge invariant. However the spatial projection of these variables defined in equation (\ref{GI1}) vanishes in the background, and so are gauge invariant and covariantly defined in the physical universe. Also it is important to define quantities which are more easy to  measure, thus is defined the fractional density gradient
\begin{equation}
 \mathcal{X}_{\alpha}=\frac{X_{\alpha}}{\kappa \mu} \quad \textrm{and} \quad \mathcal{Y}_{\alpha}=\frac{Y_{\alpha}}{\kappa P}.
\end{equation}  
In the same way one can define the gauge invariant for magnetic fields $\mathcal{B}_{\alpha}$ in a magnetized universe \cite{tsagasd}. For instance, the comoving fractional  magnetic energy density distributions and the  magnetic field vector  can be defined as follows
\numparts\begin{eqnarray}
 \mathcal{B}_{\alpha}=D_{\alpha}B^{2}, \label{invb2}\\
\mathcal{B}=\frac{a^2}{B^{2}}D^{\alpha}\mathcal{B}_{\alpha}, \label{invb21}\\
 \mathcal{M}_{\alpha \beta}=aD_{\beta}B_{\alpha},\label{invbi}
\end{eqnarray}\endnumparts
with $B^2$ the local density of the magnetic field.  As it has been argued by Tsagas et.al.  \cite{tsagasb}, they describe the spatial variation in the magnetic energy density and the spatial inhomogeneites in the distribution of the vector field $B_{i}$, as measured by a pair of neighbouring fundamental observers (which represent the motion of typical observers in the Universe being the four-velocity its vector tangent) in a gauge-invariant way.  A further discussion of fundamental observers and the meaning of these gauge invariant respect to these observers is given in section 6.3.1 of \cite{barrow}. 
\section{Gauge invariant approach}
Let us begin by reviewing some general ideas about the gauge invariant approach. 
Following \cite{bruni1,bruni2}, consider two  Lorentzian manifolds ($\mathcal{M}, g$) and ($\mathcal{M}_0, g_0$),  that represent the physical and the background space-times  respectively. The
perturbation of a tensor field $T$ is  defined as the difference between the values that the quantity takes in $\mathcal{M}$ and $\mathcal{M}_0$, evaluated at points corresponding to
the same physical event.  To compare any quantity in the two spacetimes, a  diffeomorphism $\phi:\,\mathcal{M}\rightarrow \mathcal{M}_0$  is defined which  enables the 
identification of points between  $\mathcal{M}$ and $\mathcal{M}_0$. 
However, this identification map is completely arbitrary; this freedom arises in the cosmological perturbation theory and  one may refer to it as gauge freedom of the second kind in order to distinguish it from the usual gauge freedom of general relativity \cite{sach}. Once an  identiﬁcation map $\phi$ has been assigned,  perturbations (living on $\mathcal{M}_0$) can be defined as
\begin{equation}
 \Delta^{\phi} T|_{\mathcal{M}_0}=\phi^{*}T-T_0, \label{eq1}
\end{equation}
with  $T_0$  the background tensor field corresponding to $T$ and $\phi^{*}T$  is the pull-back which gives the representation of $T$ over $\mathcal{M}_0$. 
To define the perturbation to a given order,  the fields are  expanded in a Taylor power series  and  the above mentioned iteration scheme is then used.
For this, consider a family of four-{\it sub}manifold $\mathcal{M}_\lambda$ with $\lambda \in \mathbb{R}$ embedded in a five-manifold $\mathcal{N}=\mathcal{M}\times \mathbb{R}$ \cite{nakamura1,bruni1}. Each submanifold in the family represents a perturbed space-time and the background is represented when  $\lambda=0$ (namely $\mathcal{M}_0$). In each submanifold, the Einstein and Maxwell equations must be fulfilled
\begin{equation}
 E[g_\lambda,T_\lambda]=0 \quad \textrm{and} \quad M[F_\lambda,j_\lambda]=0. 
\end{equation}
 To generalize  the definition of perturbation given in equation (\ref{eq1}), a  one-parametric group of diffeomorphisms $\mathfrak{X}_\lambda$ is introduced  in order to identify points of the background with the physical space-time labeled with $\lambda$. Therefore, one  gets a way for defining the perturbation for any tensor field
\begin{equation}
 \Delta^{\phi} T|_{\mathcal{M}_0}=\mathfrak{X}_\lambda^{*}T-T_0. \label{eq2}
\end{equation}
The first term of equation (\ref{eq2}) which lives on $\mathcal{M}_0$ admits an expansion around $\lambda=0$ given by
\begin{equation}
 \mathfrak{X}_\lambda^{*}T=\sum_{k=0}^{\infty}\frac{\lambda^k}{k!}\delta_{\mathfrak{X}}^{(k)}T=\sum_{k=0}^{\infty}\frac{\lambda^k}{k!}\mathcal{L}_{X}^{k}T|_{\mathcal{M}_0}=exp(\lambda \mathcal{L}_{X})T|_{\mathcal{M}_0},\label{eq2.5}
\end{equation}
where $\mathcal{L}_{X}T$ is the Lie derivative of $T$ along to the vector field $X$ that generates the flow $\mathfrak{X}$, $k$ does mention to the expansion order and $\delta_{\mathfrak{X}}^{(k)}T$ represents the k-th order perturbative of $T$. If we choose another vector field (gauge choice) $\mathfrak{Y}_\lambda$, the expansion of $T$ is written as
\begin{equation}
 \mathfrak{Y}_\lambda^{*}T=\sum_{k=0}^{\infty}\frac{\lambda^k}{k!}\delta_{\mathfrak{Y}}^{(k)}T=\sum_{k=0}^{\infty}\frac{\lambda^k}{k!}\mathcal{L}_{Y}^{k}T|_{\mathcal{M}_0},\label{eq2.6}
\end{equation}
At this point,  it is useful to define fields on $\mathcal{M}$ that are intrinsically gauge independent. We say that a quantity  is gauge invariant if it's value at any point of $\mathcal{M}$  does not depend on the gauge choice, namely $\mathfrak{Y}_\lambda^{*}T=\mathfrak{X}_\lambda^{*}T$. An alternative way to define a gauge invariant quantity at order $n \geqslant 1$ (see proposition 1 in \cite{bruni1}), is iff
\begin{equation}
 \mathcal{L}_{\xi} \delta^{(k)} T=0,
\end{equation}
 is satisfied. Here $\xi$ is any vector field on  $\mathcal{M}$ and $\forall k \leqslant n$. At first order ($k=1$) the Stewart-Walker lemma is found \cite{steward2}. In cases where tensor field is gauge dependent, it is useful to represent this tensor from a particular gauge  $\mathfrak{X}$ in  other  $\mathfrak{Y}$. For this, the identification map $\Phi$ on $\mathcal{M}_0$, $\Phi_\lambda:\,\mathcal{M}_0\rightarrow \mathcal{M}_0 $ is defined by
 \begin{equation}
 \Phi_\lambda=\mathfrak{X}_{-\lambda} \circ \mathfrak{Y}_\lambda \quad \textrm{that implies} \quad \mathfrak{Y}_\lambda^{*}T=\Phi^{*}_{\lambda}\mathfrak{X}_\lambda^{*}T \label{eq3}.
\end{equation}
Therefore, $\Phi$ induces a pull-back which changes the representation $X$ of $T$ to the representation $Y$ of $T$. Now, to generalize equation ({\ref{eq2.5}}) and using the Baker-Campbell-Haussdorf formula \cite{baker}, the gauge transformation on  $\mathcal{M}_0$ of $T$ is
 \begin{equation}
\Phi^{*}_{\lambda}\mathfrak{X}_\lambda^{*}T=\exp\left(\sum_{k=1}^{\infty}\frac{\lambda^k}{k!}\mathcal{L}_{\xi_k} \right) \mathfrak{X}_\lambda^{*}T,\label{eq4}
\end{equation}
with $\xi_k$ a vector field on $\mathcal{M}_\lambda$. The relations to first and second order perturbations of $T$ in two differents gauge choices are found subsituting the equations (\ref{eq2.5},\ref{eq2.6}) in equation (\ref{eq4}) obtaining
\numparts \begin{eqnarray}
\delta_{\mathfrak{Y}}^{(1)}T-\delta_{\mathfrak{X}}^{(1)}T= \mathcal{L}_{\xi_1}T|_{\mathcal{M}_0}\,,\label{gt1}\\
\delta_{\mathfrak{Y}}^{(2)}T-\delta_{\mathfrak{X}}^{(2)}T= 2\mathcal{L}_{\xi_1}\delta_{\mathfrak{X}}^{(1)}T|_{\mathcal{M}_0}+\left(\mathcal{L}_{\xi_1}^2+\mathcal{L}_{\xi_2}  \right)T|_{\mathcal{M}_0}\,.\label{gt2}
\end{eqnarray}\endnumparts
Where the generators of the gauge transformation $\Phi$ are
\begin{equation}
\xi_1=Y-X \quad \textrm{and} \quad \xi_2=[X,Y].
\end{equation}
This vector field can be split in their time and space part
\begin{equation}\label{vectoprr}
\xi_\mu^{(k)}=(\alpha^{(k)},\partial_i \beta^{(k)}+d_i^{(k)}),
\end{equation}
here  $\alpha^{(k)}$ and $ \beta^{(k)}$  are arbitrary scalar functions, and  $\partial^i d_i^{(k)}=0$. The function $\alpha^{(k)}$ determines the choice of constant time hypersurfaces, while  $ \beta^{(k)}$ and $d_i^{(k)}$ fix the spatial coordinates within the hypersurface.  
\subsection{ Perturbations on a magnetized FLRW background}\label{background}
At zero order (background), the universe is well des\-cribed by a spatially flat  FLRW
\begin{equation}
 ds^{2}=a^{2}(\tau)\left(-d\tau^{2}+ \delta_{ij}dx^{i}dx^{j}\right), 
\end{equation}
with $a(\tau)$ the scale factor with $\tau$ the conformal time. The Einstein  tensor components in this background are given by
\begin{equation}
G^{0}_{0}=-\frac{3H^{2}}{a^{2}}, \quad
 G^{i}_{j}=-\frac{1}{a^{2}}\left(2 \frac{a^{\prime\prime}}{a}-H^{2}\right)\delta^{i}_{j},
\end{equation}
with $H=\frac{a^{\prime}}{a}$ the Hubble parameter  and prime denotes the derivative with respect to  $\tau$. We consider the background  filled with a single barotropic fluid 
where  the  energy momentum tensor is 
\begin{equation}
T_{(fl)\:\nu}^{\mu}=\left(\mu_{(0)}+P_{(0)}\right)u^{\mu}_{(0)}u_{\nu}^{(0)}+P_{(0)}\delta_{\:\nu}^{\mu},\label{tflcero} 
\end{equation}
with $\mu_{(0)}$ the energy density and $P_{(0)}$ the pressure. The comoving observers are defined by the four-velocity $u^{\nu}=(a^{-1},0,0,0)$   with $u^{\nu}u_{\nu}=-1$ and the conservation law for the fluid yields 
\begin{equation}
 \mu_{(0)}^{\prime}+3H(\mu_{(0)}+P_{(0)})=0. 
\end{equation}
We also allow the presence of a weak and  spatially homogeneous  large-scale magnetic  field into our FLRW background with the property $B^{2}_{(0)}\ll \mu_{(0)}$. This field  must be sufficiently random to satisfy $\langle B_{i}^{(0)} \rangle=0$ and  $\langle B_{(0)}^{2} \rangle\neq 0$ to ensure that symmetries and the evolution of the background remains unaffected.
Working under MHD approximation  in  large scales, the plasma is globally neutral, this means that charge density is  neglected and the electric field with the current should be zero, thus the only nonzero magnetic variable in the background is  $B^{2}_{(0)}$.  The evolution of energy density magnetic field is given by
\begin{equation}
 B_{(0)}^{2 \,\prime}=-4HB_{(0)}^2, 
\end{equation}
showing $B^2 \sim a^{-4}$ in the background. 
Fixing the background,  we  consider the  perturbations up to second order about this  FLRW magnetized universe, so that  metric tensor is given by 
\begin{eqnarray}
g_{00}&=&-a^{2}(\tau)\left(1+2\psi^{(1)}+\psi^{(2)}\right),\\
g_{0i}&=&a^{2}(\tau)\left(\omega_{i}^{(1)}+\frac{1}{2}\omega_{i}^{(2)}\right),\\
g_{ij}&=&a^{2}(\tau)\left[\left(1-2\phi^{(1)}-\phi^{(2)}\right)\delta_{ij}+\chi_{ij}^{(1)}+\frac{1}{2}\chi_{ij}^{(2)}\right]. \end{eqnarray}
The perturbations are splitting into scalar, transverse vector part, and transverse trace-free tensor
\begin{equation}
\omega_{i}^{(k)}=\partial_{i}\omega^{(k)\Vert}+\omega_{i}^{(k)\bot}, \label{omega1}\end{equation}
with $\partial^{i}\omega_{i}^{(k)\bot}=0$, and $k=1,2$ \cite{bruni1}.  Similarly we can split $\chi_{ij}^{(k)}$ as
\begin{equation}
\chi_{ij}^{(k)}=D_{ij}\chi^{(k)\Vert}+\partial_{i}\chi_{j}^{(k)\bot}+\partial_{j}\chi_{i}^{(k)\bot}+\chi_{ij}^{(k)\top}, \label{chi1}
\end{equation}
for any tensor quantity\footnote{With $\partial^{i}\chi_{ij}^{(k)\top}=\partial^{i}\chi_{i}^{(k)\bot}=0$, $\chi_{i}^{(k)i}=0$  and $D_{ij}\equiv\partial_{i}\partial_{j}-\frac{1}{3}\delta_{ij}\partial_{k}\partial^{k}$.}.  
keeping in mind that  zero order the variables  depend only on $\tau$, we expand  the scalar variables such as  energy density of the matter and the magnetic field  as
\begin{eqnarray}
\mu=\mu_{(0)}+\mu_{(1)}+\frac{1}{2}\mu_{(2)},\label{f1dexpansion}\\
B^{2}=B^{2}_{(0)}+B^{2}_{(1)}+\frac{1}{2}B^{2}_{(2)},\label{f2dexpansion}
\end{eqnarray}
 and the vector variables such as magnetic and electric field and four-velocity among others as 
\begin{eqnarray}
B^{i}=\frac{1}{a^{2}(\tau)}\left(B^i_{(1)}+\frac{1}{2}B_{(2)}^{i}\right),\label{Bexpansion}\\
E^{i}=\frac{1}{a^{2}(\tau)}\left(E^i_{(1)}+\frac{1}{2}E_{(2)}^{i}\right),\label{Bexpansion}\\
u^{\mu}=\frac{1}{a(\tau)}\left(\delta_{0}^{\mu}+v_{(1)}^\mu+\frac{1}{2}v_{(2)}^{\mu}\right).
\end{eqnarray}
Again, the 4-velocity $u^\mu$ is subject to normalization condition $u^\mu u_\mu=-1$, and in any gauge it can be expressed as
 \begin{eqnarray}
u_{\mu}=a \left[-1-\psi^{(1)}-\frac{1}{2}\psi^{(2)}+\frac{1}{2}\psi^{2}_{(1)}-v^{(1)}_{i}v_{(1)}^{i},\right. \nonumber \\
  \left.\omega^{(1)}_{i}+v^{(1)}_{i} +\frac{1}{2}\left(\omega^{(2)}_{i}+v^{(2)}_{i}\right)-\omega^{(1)}_{i}\psi^{(1)}+v^{j}_{(1)}\chi^{(1)}_{ij}-2v_{i}^{(1)}\phi_{(1)} \right]\label{u2cov}
\end{eqnarray}
\begin{equation}
 u^{\mu}=\frac{1}{a}\left[1-\psi^{(1)}+\frac{1}{2}\left(3\psi^{2}_{(1)}-\psi^{(2)}+v^{(1)}_{i}v_{(1)}^{i}+2\omega_{i}^{(1)}v^{i}_{(1)}\right),v^{i}_{(1)}+\frac{1}{2}v^{i}_{(2)}\right].\label{u2cont}
\end{equation}
With the 4-velocity one can also define the aceleration as
\begin{equation}
 a_{\mu}=u^{\nu}\nabla_\nu u_{\mu}.\label{4acel}
\end{equation}
Using the  equation (\ref{gt1}), we can find the transformation of the  metric and matter variables at first order
\numparts\begin{eqnarray}
\tilde{\psi}^{(1)}=\psi^{(1)}+\frac{1}{a}(a\alpha_{(1)})^\prime,\label{pot2a}\\
\tilde{\phi}^{(1)}=\phi^{(1)}-H\alpha_{(1)}-\frac{1}{3}\nabla^2\beta^{(1)},\label{pot2b}\\
 \tilde{v_{i}}^{(1)}=v_{i}^{(1)}-\xi^{\prime\,(1)}_{i},\label{pot2c}\\
\tilde{\omega_{i}}^{(1)}=\omega_{i}^{(1)}-\partial_{i}\alpha^{(1)}+\xi^{\prime}_{i\,(1)},\label{pot2d}\\
\tilde{\chi}^{(1)}_{ij}=\chi_{ij}^{(1)}+\partial_{i}\xi^{(1)}_j+\partial_{j}\xi^{(1)}_i-\frac{2}{3}\delta_{i j}\nabla^2\beta,\label{pot2e}
\end{eqnarray}\endnumparts
and with these latter equations, we can build the gauge invariant variables. One way for getting the gauge invariant,  is to fix the  vector field  $\xi$ at a particular gauge, for  example the longitudinal gauge (set the scalar perturbations $\omega$ and $\chi$ being zero). So, the scalar \textit{gauge invariant variables} at first order are given by
\begin{equation}
\Psi^{(1)}\equiv\psi^{(1)}+\frac{1}{a}\left(\mathcal{S}_{(1)}^{||}a\right)^{\prime}, \quad \textrm{and} \quad \Phi^{(1)}\equiv\phi^{(1)}+\frac{1}{6}\nabla^{2}\chi^{(1)}-H\mathcal{S}_{(1)}^{||},\\
\end{equation}
 with $\mathcal{S}_{(1)}^{||}\equiv\left(\omega^{||(1)}-\frac{\left(\chi^{||(1)}\right)^{\prime}}{2}\right)$ the scalar contribution of the shear. These are commonly called the Bardeen potentials
which were interpreted by Bardeen  as the spatial dependence of the proper time intervals between two nearly observers and curvature perturbations respectively \cite{bardeen}. Other scalar invariants are   
\begin{equation}
\Delta^{(1)}\equiv\mu_{(1)}+\left(\mu_{(0)}\right)^{\prime}\mathcal{S}_{(1)}^{||}, \quad \textrm{and}\quad \Delta^{(1)}_{P}\equiv P_{(1)}+\left(P_{(0)}\right)^{\prime}\mathcal{S}_{(1)}^{||},
\end{equation}
 which  describe  the energy density and  pressure of the matter. The vector modes yields
\begin{equation}
\vartheta_{i}^{(1)}\equiv\omega_{i}^{(1)}-\left(\chi_{i}^{\bot(1)}\right)^{\prime},\quad \textrm{and} \quad \mathcal{V}_{(1)}^{i}\equiv\omega_{(1)}^{i}+v_{(1)}^{i},
\end{equation}
related  with the vorticity of the fluid.  There are other  gauge invariant variables at first order such as   the 3-current,  the charge density and the electric and magnetic fields, because they vanish in the background. Tensor quantities are also gauge invariant because they are null in the background \cite{steward2}.
In order to study the evolution of magnetic field in large-scales we must rewrite  Maxwell's  equation (\ref{maxwelldef}) in this formalism. The deduction of the following equations  is  shown in \cite{hortua}. At first order the Maxwell's equation are expressed as
\numparts\begin{eqnarray}
\partial_{i}E^i_{(1)}=a\varrho_{(1)}\,,\\
\partial_{i}B^i_{(1)}=0\,, \\
\epsilon^{ilk}\partial_l B_k^{(1)}=(E^i_{(1)})^\prime+2HE^i_{(1)}+aJ^i_{(1)}\,,\\
(B^i_{(1)})^\prime+2HB^i_{(1)}=-\epsilon^{ilk}\partial_l E_k^{(1)}\,,
\end{eqnarray}\endnumparts
these equations represent the evolution of  fields in a totally invariant way. Furthermore, the energy density of the magnetic field is the unique variable which is gauge dependent  and evolves  under MHD approximation as $\sim a^{-4}$ and  transforms at first order as
\begin{equation}
\tilde{B}^{2}_{(1)}=B^{2}_{(1)}+\left(B^{2}_{(0)}\right)^{\prime}\alpha^{(1)}. \label{energymf1}
\end{equation}
At second order,  the  Maxwell's equations are given by \cite{hortua}
\numparts\begin{eqnarray}
\partial_{i}E_{(2)}^{i}=-4E_{(1)}^{i}\partial_{i}\left(\psi^{(1)}-3\phi^{(1)}\right)+a\varrho^{(2)},\label{maxwell2a}\\
\left(\nabla\times B^{(2)}\right)^{i}=2E_{(1)}^{i}\left(2\left(\psi^{(1)}\right)^{\prime}-6\left(\phi^{(1)}\right)^{\prime}\right)+\left(E_{(2)}^{i}\right)^{\prime}+2HE_{(2)}^{i}\nonumber\\
+2\left(\nabla\left(2\psi^{(1)}-6\phi^{(1)}\right)\times B_{(1)}\right)^{i}+aJ_{(2)}^{i},\label{maxwell2b}\\
 \frac{1}{a^{2}}\left(a^{2} B_{k}^{(2)}\right)^{\prime}+\left(\nabla\times E_{j(2)}\right)_{k}=0\,, \label{maxwell2c}\\
\partial_{i}B^{i(2)}= 0\,\label{maxwell2d}. 
\end{eqnarray}\endnumparts
dependent on gauge choice. The magnetic gauge dependent variables  transform as 
\begin{eqnarray}
\tilde{E}^{(2)}_{i} = E^{(2)}_{i}+2\left[\frac{\left(a^{2}E^{(1)}_{i} \alpha^{(1)}\right)^{\prime}}{a^{2}}+\left(\xi^{\prime}_{(1)}\times B^{(1)}\right)_{i}+ \xi^{l}_{(1)}\partial_{l}E_{i}^{(1)}+E_{l}^{(1)}\partial_{i}\xi^{l}_{(1)}\right],\label{etrans} \\
\tilde{B}^{(2)}_{i} = B^{(2)}_{i}+2\left[\frac{\alpha^{(1)}}{a^{2}}\left(a^{2}B^{(1)}_{i} \right)^{\prime}+\left(\nabla \times \left(B^{(1)}\times\xi^{(1)}\right)+E^{(1)}\times \nabla\alpha^{(1)}\right)_{i}\right], \label{btrans}\end{eqnarray}
here $\varrho^{(2)}$ and $J_{(2)}^{i}$ transform in according to  equations (80),(81) in \cite{hortua}. The energy density at second order evolves as equation (117) in \cite{hortua}  and  it transforms 
\begin{eqnarray}
\tilde{B}^2_{(2)}=B^{2}_{(2)} +B^{2\prime}_{(0)} \alpha_{(2)}+\alpha_{(1)}\left(B^{2\prime \prime}_{(0)}\alpha_{(1)}+B^{2\prime}_{(0)}\alpha_{(1)}^\prime+2B^{2\prime}_{(1)}\right)\nonumber\\
+\xi^i_{(1)}\left(B^{2\prime}_{(0)}\partial_{i}\alpha^{(1)}
+2\partial_{i}B^{2}_{(1)}\right). \label{ultimm}
\end{eqnarray}
Fixing the gauge  we find out gauge invariant variables related with the electromagnetic fields. Finally, applying the divergence to  equation (\ref{maxwell2b}) and using the  equation (\ref{maxwell2a}), we obtain the conservation´s equations up to second order  for the charge given by
\begin{equation}
\varrho^{\prime}+3H\varrho+\nabla \cdot J=0.\label{eqconser2}\end{equation} 
Here, at first order approximation, the equation is completly invariant, but at second order the involved variables are gauge dependent and transform according to (80),(81) in \cite{hortua}.  

\section{Electromagnetic potentials}\label{sectionepo}
 In order to  study  the behavior of electromagnetic fields in scenarios such as  inflation, vector-tensor theories  \cite{jimenez2,jimenez3} or quantization of gauge theories in nontrivial spacetimes \cite{jimenez1}, it is  more convenient to write the Maxwell's equations in terms of a four-potential. Therefore, in this section we will apply the gauge invariant approach to scenarios where the prensence electromagnetic four-potential becomes  relevant. 
The covariant form of the Maxwell's equations (see homogeneous equation (\ref{maxwelldef}))  reflects  the existence of a four-potential \cite{tsagasb}. This means, we can  define the four potential as $A_{\mu}=(-\varphi,A_i)$ with the antisymmetric condition given by $F_{\mu \nu}=\partial_\nu A_\mu - \partial_\mu A_\nu$. At first order, the four-potential is gauge invariant (because they are  null at the background)\footnote{The magnetic potential is null at the background  while that electric potential at most is a constant, but due to Stewart-Walker lemma it is gauge invariant. }. Using the homogeneous Maxwell´s equations , we can define the fields in terms of four-vector potentials
\begin{equation}
B^{(1)}_i=(\nabla \times A^{(1)})_i \quad \textrm{and} \quad E_{i}^{(1)}=-(A_i^{(1)\,\prime}+2HA_i^{(1)}+\partial_i \varphi^{(1)}). 
\end{equation}
 Therefore the inhomogeneous Maxwell's  equations could be   reduced to two invariant equations
\begin{eqnarray}
\nabla^2 \varphi^{(1)} +\frac{1}{a^2}\frac{\partial}{\partial t}\left(\nabla \cdot (a^2 A^{(1)})\right)=-a\varrho_{(1)}\\
\nabla^2 A^{(1)}_i-\frac{1}{a^2}\frac{\partial^2}{\partial t^2}(a^2 A^{(1)}_i)-\partial_i\left(\nabla \cdot  A^{(1)}+\frac{1}{a^2}\frac{\partial }{\partial t}(a^2\varphi^{(1)})\right)=-aJ_i^{(1)}.
\end{eqnarray}
The latter equations although are written in terms of gauge invariant quantities, they have an arbitrariness in the potentials known in electrodynamics given by the transformations  $\widetilde{A}^{(1)}_i=A^{(1)}_i+\partial_i\Lambda$ and $\widetilde{\varphi}^{(1)}=\varphi^{(1)}-\frac{1}{a^2}\frac{\partial}{\partial t}( a^2\Lambda)$, being $\Lambda$ some scalar function of same order that potentials and  where the fields are left unchange under this transformation. It is commonly known in the literature, the freedom given by this transformation implies we can choose the  set of potentials satisfy the Lorenz conditions which in this case is 
\begin{equation}
\nabla \cdot  A^{(1)}+\frac{1}{a^2}\frac{\partial }{\partial t}(a^2\varphi^{(1)})=0.
\end{equation}
Therefore, we can arrive to an uncoupled set of equations for the potentials, which are equivalents to Maxwell equations
\begin{eqnarray}
\nabla^2 \varphi^{(1)} -\frac{1}{a^2}\frac{\partial^2}{\partial t^2}(a^2\varphi^{(1)})=-a\varrho_{(1)}\\
\nabla^2 A^{(1)}_i-\frac{1}{a^2}\frac{\partial^2}{\partial t^2}(a^2 A^{(1)}_i)=-aJ_i^{(1)}.
\end{eqnarray}
At second order the procedure is more complex given the gauge dependence of the potentials. Using the antisymmetrization and the gauge transformation equation (\ref{gt2}),  we have found that the four-portential
transforms as 
\begin{eqnarray}
\tilde{\varphi}^{(2)} = \varphi^{(2)}+2\left[\frac{\alpha_{(1)}}{a^2}(a^2\varphi_{(1)})^\prime+\xi^{i}_{(1)}\partial_i\varphi^{(1)}+\alpha_{(1)}^\prime\varphi_{(1)}-\xi_i^{(1)\,\prime}A^{i}_{(1)} \right],\label{phitrans} \\
\tilde{A}^{(2)}_{i} = A^{(2)}_{i}+2\left[\frac{\alpha_{(1)}}{a^2}(a^2 A^{(1)}_i)^\prime+\partial_l A_i^{(1)}\xi^l_{(1)}-\varphi_{(1)}\partial_i\alpha^{(1)}+A^l_{(1)}\partial_i \xi_l^{(1)}  \right].\label{Atrans} \end{eqnarray}
Applying the curl operator at vector potential  $\mathcal{A}^{(2)}_{i}$ and  after a long but otherwise straightforward algebra,  we obtain the transformation of magnetic field given by equation (\ref{btrans}) and the vector potential can expressed as
\begin{equation} 
\tilde{B}^{(2)}_{i}=(\nabla \times \tilde{A}^{(2)})_{i},\label{A2order}
\end{equation}
being a original result of this paper.  Similarly, we can use the induction equation (\ref{maxwell2b}) found in the previous  section, and with some algebra we found that the scalar potential is described in terms of electric field equation (\ref{etrans}) via
\begin{equation}
\partial_i \tilde{\varphi}^{(2)}=-\tilde{E}^{(2)}_{i}-\frac{1}{a^2}\left(a^2\tilde{A}_{i}^{(2)}\right)^{\prime}, \label{phi2order} 
\end{equation}
again the four-potential at this order has a freedom mediated by some scalar function $\Lambda$ with same order and under similar transformations showed at first order, the fields $E^{(2)}_{i}$ and $B^{(2)}_{i}$ are left unchanged. Let us continue with the Maxwell equation at second order written in terms of the four-potential. For this purpose, we substitute the equations (\ref{A2order}),(\ref{phi2order}) in the inhomogeneous Maxwell equations (\ref{maxwell2a}), (\ref{maxwell2b}) obtaining  a coupling set of equations given by   
\begin{eqnarray}
\nabla^2 \varphi^{(2)} +\frac{1}{a^2}\frac{\partial}{\partial t}\left(\nabla \cdot (a^2 A^{(2)})\right)-4\left(\frac{1}{a^2}(a^2A_i^{(1)})^\prime+\partial_i\varphi^{(1)}\right)\times \nonumber \\
\partial^i(\psi^{(1)}-3\phi^{(1)})=-a\varrho^{(2)},\\
\nabla^2 A^{(2)}_i-\frac{1}{a^2}\frac{\partial^2}{\partial t^2}(a^2 A^{(2)}_i)-\partial_i\left(\nabla \cdot  A^{(2)}+\frac{1}{a^2}\frac{\partial}{\partial t}(a^2\varphi^{(2)})\right)\nonumber\\
-4\left(\frac{1}{a^2}(a^2A_i^{(1)})^\prime+\partial_i\varphi^{(1)}\right)\left(\psi_{(1)}^\prime-3\phi_{(1)}^{\prime}\right)+4\left(\nabla^2 A^{(1)}_i-\partial_i(\nabla \cdot A^{(1)})\right)\times \nonumber\\
\left(\psi_{(1)}-3\phi_{(1)}\right)=-aJ_i^{(2)}.
\end{eqnarray}
in a dependent gauge way. The gravitational potentials $\psi$ and $\phi$ transform  via the  equations (\ref{pot2a}), (\ref{pot2b}). With these equations we can see a strong dependence between the electromagnetic fields and the gravitational effects with  first order couplings  between these variables.
The Maxwell´s  equation found above, are still gauge dependent   due to the fact that  electromagnetic and gravitational potentials have a freedom in the choice of  $\xi^{\nu}$,  the gauge vector.  Thus fixing the value of $\xi^{\nu}$,  the  variables might  take on their given meaning.  For example, assuming that
\begin{equation}
 \tilde{\psi}^{(1)}-3\tilde{\phi}^{(1)}=0, \label{gaugephi} 
\end{equation}
in order to have the  same expression gotten in the first order case, and  using the equations (\ref{pot2a})-(\ref{pot2b}),  an important constraint for the vector part of the gauge dependence is found
\begin{equation}
-\nabla^2\beta^{(1)}=\psi^{(1)}-3\phi^{(1)}+4H\alpha^{(1)}+\alpha_{(1)}^{\prime}.\label{fix2b}
\end{equation}
With this choice, the conservation´s equation given by expression (\ref{eqconser2}) reads as
\begin{equation}
\Delta_{\varrho}^{(2)\prime}+3H\Delta_{\varrho}^{(2)}+\partial_i \mathcal{J}_{(2)}^{i} \nonumber\\
+2\varrho_{(1)}(\Psi_{(1)}^{\prime}-3\Phi_{(1)}^{\prime})+2J_{(1)}^i\partial_i(\Psi_{(1)}-3\Phi_{(1)})=0,\label{eqconser3} \end{equation}
which is gauge invariant and equivalent  to the equation (B2) in \cite{hortua}.
We can also  use the Lorenz condition by fixing the freedom of the fields
\begin{equation}
\nabla \cdot  A^{(2)}+\frac{1}{a^2}\frac{\partial }{\partial t}(a^2 \varphi^{(2)})=0,
\end{equation}
obtaining the Maxwell´s equation in terms of the potential and written in a invariant way.   
\section{Equivalence between two approaches}
In this section we present the method to find the equivalence between  both approaches mentioned above. For doing this, we  compare the  gauge invariant quantities built in each approach similar to that   used  by \cite{bruni2} and \cite{malik3}. The comoving gauge is defined by choosing spatial coordinates such that the $3-$velocity of the fluid vanishes  $\tilde{u}^{i}=0$, and the four-velocity is orthogonal to hypersurface of constant time \cite{malik2}.
From equation (\ref{u2cov})  we have $\tilde{\omega_{i}}^{(1)}+\tilde{v_{i}}^{(1)}=0$ and  using the equations (\ref{pot2c}), (\ref{pot2d}) we fix the values for the gauge transformation generator vector field $\xi^\mu$ given by
\begin{eqnarray}\label{gaugedep1}
 \tilde{\omega_{i}}^{(1)}+\tilde{v_{i}}^{(1)}=0 \rightarrow \alpha^{(1)}=v^{\parallel}+\omega^{\parallel}\,, \label{alpha111} \nonumber \\
 \tilde{v}^{\parallel(1)}=0 \rightarrow \beta^{(1)}=\int v^{\parallel}d\tau +C^{\parallel}(x^{i}),\nonumber \\
 \tilde{v_{i}}^{(1)}=0 \rightarrow d^{(1)}_{i}=\int v^{\perp}_{i}d\tau +C^{\perp}_{i}(x^{i}),
\end{eqnarray}
with $C(x^i)$ a residual gauge freedom.   Therefore by using this constraint for  $\xi^\mu$ (see equation (\ref{vectoprr})), we can define a gauge invariant quantity related with the energy density of the magnetic field in the gauge invariant approach by substituting the value of $\xi^\mu$  from (\ref{alpha111}) in  equation (\ref{energymf1})  obtaining 
 \begin{equation}\label{giq}
\Delta^{(1)}_{mag}:=\tilde{B}^{2}_{(1)}=B^{2}_{(1)}+\left(B^{2}_{(0)}\right)^{\prime}(v^{\parallel}_{(1)}+\omega^{\parallel}_{(1)}),\quad \rightarrow\textbf{Comoving Gauge.}
\end{equation}
 Now, we start  expanding  the equation  (\ref{invb2}), where we use the projector defined in equation (\ref{oper}) and the four-velocity given by equation  (\ref{u2cov}); at first order we obtain
\begin{equation}
\mathcal{B}_{0}= D_{0}B^{2}_{(1)}=0,\end{equation}
for the temporal part. For spatial part we get
\begin{equation}\label{eqqq1}
\mathcal{B}_{i} = D_{i}B^{2}_{(1)}=\partial_{i}\left(B^{2}_{(1)}+\left(B^{2}_{(0)}\right)^{\prime}\left(v^{\parallel}_{(1)}+\omega^{\parallel}_{(1)}\right)\right),
\end{equation}
where both equations correspond to the gauge invariant in the $1+3$-covariant approach.  If we compare the latter equation with the gauge invariant quantity corresponding to  energy density of magnetic field (see equation(\ref{giq})), we have finally 
\begin{equation}\label{equvalence1}
 \mathcal{B}_{i}= D_{i}B^{2}_{(1)} \equiv \partial_{i} \Delta^{(1)}_{mag}.
\end{equation}
The authors in \cite{malik3} found  similar results for the matter density case.
For describing the equivalence at second order, we  will make use  $\tilde{u_{i}}=0$ again (comoving condition), thus checking the equation (\ref{u2cov}) we found that
\begin{equation}
 \frac{1}{2}\left(\tilde{\omega}^{(2)}_{i}+\tilde{v}^{(2)}_{i}\right)-\tilde{\omega}^{(1)}_{i}\tilde{\psi}^{(1)}-2\tilde{v}^{(1)}_{i}\tilde{\phi}^{(1)}+\tilde{v}_{(1)}^{j}\tilde{\chi}^{(1)}_{ij}=0.\label{cond2a}
\end{equation}
 Substituting  equations (\ref{pot2a})-(\ref{pot2e}) and values for $\tilde{\omega}^{(2)}_i$, $\tilde{v}^{(2)}_i$, and $\tilde{\chi}^{(1)}_{ij}$   in the latter equation,   we obtain the temporal gauge dependence  $\alpha^{(2)}$ written in the comoving gauge given by
\begin{eqnarray}\label{eq:al2}
\partial_{i}\alpha^{(2)}&=&\omega_{i}^{(2)}+v_{i}^{(2)}-4\psi^{(1)}\left(\omega_{i}^{(1)}+v_{i}^{(1)}\right)+2v_{i}^{(1)}\left(\psi^{(1)}-2\phi^{(1)}\right) \nonumber\\
&+&\left(\omega_{\parallel}^{(1)}+v^{(1)}_{\parallel}\right)\left(\omega_{i}^{(1)}+v_{i}^{(1)} \right)^{\prime}-\left(\omega_{\parallel}^{(1)}+v^{(1)}_{\parallel}\right)^{\prime}\left(\omega_{i}^{(1)}+v_{i}^{(1)} \right) \nonumber\\
&+&\partial_{i}\xi_{j}^{(1)}\left(\omega_{(1)}^{j}+v^{j}_{(1)}\right)+2\chi_{ij}v^{j}+\xi^{j}_{(1)}\partial_{j}\left(\omega_{i}^{(1)}+v^{(1)}_{i}\right),
\end{eqnarray}
the deduction of this equation is given in \ref{apena}. We can also define a gauge invariant quantity related with the energy density of the magnetic field in the gauge invariant approach at second order fixing  the value of $\alpha^{(2)}$  from (\ref{eq:al2}) and $\xi^{(1)}_i$  from (\ref{alpha111}) in  the equation (\ref{ultimm})  yields
 \begin{equation}\label{giq2}
\Delta^{(2)}_{mag}:=\tilde{B}^{2}_{(2)},\quad \rightarrow\textbf{Comoving Gauge.}
\end{equation}
On the other hand,  expanding the equation (\ref{invb2})  at second order (which comes from 1+3 covariant approach), the temporal part corresponds to
\begin{equation}\label{constraint1}
\mathcal{B}_{0}=D_{0}B^{2}_{(2)}=-v^{i}_{(1)}B_{(0)}^{2\prime}\left(v_{i}^{(1)}+\omega_{i}^{(1)}\right)-v^{i}_{(1)}\partial_{i} B^{2}_{(1)},
\end{equation}
where is the same result found in (\ref{eqqq1}) times $v^{i}_{(1)}$, therefore the temporal part is zero and give us an important constraint for our work.
For the spatial part we found out the following
\begin{eqnarray}\label{constraint2}
\mathcal{B}_{i}=D_{i}B^{2}=\frac{1}{2}\partial_{i}B^{2}_{(2)}+\left ( \omega_{i}^{(1)}+v_{i}^{(1)} \right) B_{(1)}^{2\prime} +B^{2\prime}_{(0)}\left(\frac{1}{2}\left(\omega_{i}^{(2)}+v_{i}^{(2)}\right) \right. \nonumber \\
- \left. 2\omega_{i}^{(1)}\psi_{(1)}-2v_{i}^{(1)}\phi_{(1)}-\psi^{(1)}v_{i}^{(1)}+\chi_{ij}^{(1)}v^{j}_{(1)}\right)
\end{eqnarray}
 Now, applying the  gradient operator $\partial_{i}$ to $\Delta^{(2)}_{(mag)}$ showed in equation (\ref{giq2}), which is  an  invariant quantity  associated with energy density at second order,  we get
\begin{eqnarray}
\partial_{i}\Delta^{(2)}_{(mag)} =\partial_{i}B_{(2)}^{2}+\partial_{i}\alpha^{(2)}B^{2\prime}_{(0)}+2\alpha_{(1)}\partial_{i}\alpha^{(1)}B^{2\prime\prime}_{(0)}+B^{2\prime}_{(0)}\left(\alpha^{(1)\prime}\partial_{i}\alpha^{(1)} +\alpha^{(1)}\partial_{i}\alpha^{\prime}_{(1)}\right)\nonumber \\
+2B^{2\prime}_{(1)}\partial_{i}\alpha^{(1)}+2\alpha^{(1)}\partial_{i}B^{2\prime}_{(1)}+\partial_{i}\xi^{j}_{(1)}\partial_{j}\alpha^{(1)}B^{2\prime}_{(0)}
+\xi^{j}_{(1)}\partial_{i}\partial_{j}\alpha^{(1)}B^{2\prime}_{(0)}\nonumber\\
+2\partial_{i}\xi^{j}_{(1)}\partial_{j}B^{2\prime}_{(1)}+2\xi^{j}_{(1)}\partial_{j}\partial_{i}B^{2}_{(1)}
\end{eqnarray}
Thus, substituting the equations (\ref{constraint1}) and (\ref{gaugedep1}) in the latter equation, we obtain
\begin{eqnarray}
\partial_{i}\Delta^{(2)}_{(mag)}=\frac{1}{2}\partial_{i}B^{2}_{(2)}+\left ( \omega_{i}^{(1)}+v_{i}^{(1)} \right) B_{(1)}^{2\prime} +B^{2\prime}_{(0)}\left(\frac{1}{2}\left(\omega_{i}^{(2)}+v_{i}^{(2)}\right) \right. \nonumber \\- \left. 2\omega_{i}^{(1)}\psi_{(1)}-2v_{i}^{(1)}\phi_{(1)}-\psi^{(1)}v_{i}^{(1)}+\chi_{ij}^{(1)}v^{j}_{(1)}
 \right),
\end{eqnarray}
which is the expression found in equation (\ref{constraint2}). Therefore we have obtained the desired result, an equivalence between the invariants of the two approaches up to second order
\begin{equation}\label{equvalence2}
\mathcal{B}_{i}=D_{i}B^{2}\equiv \partial_{i}\Delta_{mag}^{(2)}.
\end{equation}
For the gauge invariant vector field defined in equation (\ref{invbi}) we have
\numparts\begin{eqnarray}
 \mathcal{M}_{0 \,0}=0.\label{vectm0}\\
 \mathcal{M}_{[0\,i]}=\big(aB_{i(2)}\big)^{\prime}+av^j_{(1)}\partial_{[j}B^{(1)}_{i]}.\label{vectm0i}\\
 \mathcal{M}_{[i\,j]}=a\bigg(\partial_{[j} B^{(2)}_{i]}+B^{(1)\,\prime}_{[i} \mathcal{V}^{(1)}_{j]} \bigg).\label{vectmij}\\
\mathcal{M}_{i}^{\,\,\,i}=a\bigg(\partial_{i} B_{(2)}^{i} -\frac{1}{a} B_{(1)}^{i}(a\mathcal{V}_{i\,(1)})^{\prime} -3B^i_{(1)} \partial_i \phi\bigg) .\label{vectmii}
\end{eqnarray}\endnumparts
If we consider neither the magnetic field  nor  vorticity   in linear perturbation theory in equation (\ref{vectmii}), we get the usual equation of divergence of the magnetic field (which confirms a claim in \cite{Adam}). Making the antisymmetric product between  the 4-acceleration equation (\ref{4acel}) with the magnetic field, gives an equation 
of the type  
\begin{equation}
a^{(1)}_{[i}B^{(1)}_{j]}=B^{(1)}_{[i}\mathcal{V}^{(1)\,\prime}_{j]}+B^{(1)}_{[i}\partial_{j]}\psi^{(1)}+H B^{(1)}_{[i}\mathcal{V}_{j]}^{(1)},\label{4-a} 
\end{equation}
where we use the 4-velocity expressed in equation (\ref{u2cov}) from section \ref{background} and where the  temporal part is zero. If we contract the indices in the equation (\ref{4-a}) and we use the equation (\ref{gaugephi}), we get a consistency condition with the equation (\ref{vectmii}) under a null electric field condition. Therefore a  magnetic field  with no accompanying electric field and currents provides the relation
\begin{equation}
    a^{(1)}_{[\alpha}B^{(1)}_{\beta]}=\mathcal{M}_{[\alpha\,\beta]},
\end{equation}
establishing an important relation between the  gradient of the magnetic field with a  kinematic quantity as it has been argued by \cite{tsagasb}. Taking the curl of equation (\ref{btrans}) and  using the Maxwell's equation (\ref{maxwell2b}), we  find out that
\begin{equation}
(\nabla \times \tilde{H}^{(2)})_i \equiv (\nabla \times \tilde{B}^{(2)})_i=a(\nabla \times B^{(2)})_i.\label{vecta1}
\end{equation}
 where  the electric field and vorticity (this assumption  will be reflected as $\epsilon^{kij}\partial_i\xi^{(1)}_{j}=0$) have been ignored. 
Here $\tilde{H}_i^{(2)}$ is the gauge invariant quantity  related to the magnetic field vector in the gauge invariant approach. 
Therefore, by means of equations (\ref{vectmij}) and (\ref{vecta1}) allows us to find the vector equivalence up to second order given as
  \begin{equation}
\epsilon^{kij}\mathcal{M}_{[i\,j]}=(\nabla \times \tilde{H})^k,\label{vecta11}
\end{equation}
which can be described as the variations of the magnetic field vector. In short, assuming a magnetized universe  we have verified the equivalence of  both approaches by finding connections between their gauge invariant quantities   via equations (\ref{equvalence1}), (\ref{equvalence2}) for scalar   and (\ref{vecta11}) for the tensor case. 
\section{Discussion}
Relativistic perturbation theory has been an important tool in theoretical cosmology to link  scenarios of the early universe with  cosmological data such as CMB-fluctuations. However, there is an  issue in the treatment of this theory, which is called gauge problem. Due to the general covariance, a gauge degree of freedom, arises in cosmological  perturbations theory. If the correspondence between a real and background space-time is not completely specified, the evolution of the variables will have unphysical modes. Different approaches have been developed to overcome this problem, amoung them, 1+3 covariant gauge invariant and the gauge invariant approaches, which were studied in the present paper.  Following some results shown in   \cite{bruni2,bruni3,baker} and \cite{malik3}, we have contrasted these formalisms  comparing their gauge invariant variables defined in each case. Using a magnetic scenario, we have shown a strong relation between both formalisms, indeed, we found that gauge invariant defined by 1+3 covariant approach  is related with  spatial variations of
the magnetic field energy density (variable defined in the invariant gauge formalism) between two closed fundamental observers as  it is noticed in equations (\ref{equvalence1}),  (\ref{equvalence2}) and (\ref{vecta11}).  Moreover, we have also derived the gauge transformations  for electromagnetic potentials, equations (\ref{phitrans}) and (\ref{Atrans}), which are relevant in the study of evolution of primordial magnetic fields in  scenarios such as inflation or  later phase transitions. With the description of the electromagnetic potentials, we have expressed the  Maxwell's equations in terms of these ones, finding again an important coupling with the gravitational potentials.

\section{Acknowledgements}
We thank Adam J. Christopherson for interesting discussions and suggestions. The  package  VEST (Vector Einstein Summation Tools)\cite{jonathan} was used to obtain some properties in the  potentials in Section \ref{sectionepo} with the derivation of  equation (\ref{vecta11}).
\appendix
\section{Spatial part of the gauge transformation generator}\label{apena}
In order to get the equation (\ref{eq:al2}), we use the expression (\ref{gt2}) to find the way that $v^{(2)}_{i}$ and $\omega^{(2)}_{i}$ transform at second order. 
The expression was obtained first by \cite{bruni1} and becomes
\begin{eqnarray}
  \tilde{\omega}^{(2)}_i & = & \omega^{(2)}_i-\partial_i\alpha^{(2)} +\xi^{(2)\prime}_{i}+ \xi^j_{(1)}
\left(2\partial_j\omega^{(1)}_{i}-\partial_i\partial_j\alpha^{(1)}
+\partial_j\xi^{(1)\prime}_{i}\right) \nonumber\\
 &+&
\alpha^{(1)}\left[
2\left(\omega^{(1)\prime}_i +2H\omega^{(1)}_i\right)
-\partial_i\alpha^{(1)\prime} + \xi^{(1)\prime\prime}_{i} - 4H\left(\partial_i\alpha^{(1)}
 -\xi^{(1)\prime}_i\right)\right] \nonumber\\
 &+&\alpha_{(1)}^{\prime}\left( 2\omega^{(1)}_i
-3\partial_i\alpha^{(1)}  
+\xi^{(1)\prime}_{i}\right)
+ \xi^{j\prime}_{(1)}\left(-4\phi^{(1)}\delta_{ij}+2\chi^{(1)}_{ij}
+2\xi^{(1)}_{j,i}+\xi^{(1)}_{i,j}\right)\nonumber\\
&+&\xi^j_{(1),i}\left(2\omega^{(1)}_j-\partial_j\alpha^{(1)}\right)-4\psi^{(1)}\partial_i\alpha^{(1)}, \label{w2}
\end{eqnarray}
for the shear part and
\begin{eqnarray}
 \tilde{v}_i^{(2)} & = & v_i^{(2)} 
 -\xi_{i}^{(2)\,\prime}
+\alpha_{(1)}\left[
2\left(v^{(1)\,\prime}_{i}-Hv_i^{(1)}\right)
-\left(\xi^{(1)\,\prime\prime}_{i}
-2H\xi^{(1)\,\prime}_{i}\right)\right]\nonumber \\
  &+ & \xi^j_{(1)}\partial_j\left(2v^{(1)}_{i}-\xi^{(1)\,\prime}_{i}\right)
-\partial_j\xi^{(1)}_{i} \left( 2 v^j_{(1)} -\xi^{j\prime}_{(1)}\right)
+\xi^{(1)\,\prime}_{i} 
\left( 2 \psi_{(1)} +\alpha_{(1)}^\prime\right). \label{vi2}
\end{eqnarray}
for the 3-velocity. The couplings between first order terms are then given as
\begin{eqnarray}
\quad\quad\quad\quad\quad\tilde{\omega}_i^{(1)}\tilde{\psi}^{(1)}&=&(\omega_{i}^{(1)}-\partial_{i}\alpha^{(1)}+\xi^{\prime}_{i\,(1)})(\psi^{(1)}+\frac{1}{a}(a\alpha_{(1)})^\prime),\nonumber\\
-2\tilde{v}_i^{(1)}\tilde{\phi}^{(1)}+\tilde{\omega}^j_{(1)}\tilde{\chi}_{ij}^{(1)}&=&-2(v_{i}^{(1)}-\xi^{\prime\,(1)}_{i})(\phi^{(1)}-H\alpha_{(1)})\nonumber\\
&+&(v^{j\,(1)}-\xi^{j\,\prime}_{(1)})(\chi_{ij}^{(1)}+\partial_{(i}\xi^{(1)}_{j)}).\label{vi3}
\end{eqnarray}
By substituting the equations (\ref{w2}), (\ref{vi2}) and (\ref{vi3}) in equation (\ref{cond2a}) we arrive at  equation (\ref{eq:al2}). As an alternative way, we can use the equation (A12) in \cite{hortua} and transforms it from Poisson  to comoving gauge.

\end{document}